\documentclass[pdflatex,sn-mathphys-ay]{sn-jnl}

\usepackage{graphicx}%
\usepackage{multirow}%
\usepackage{amsmath,amssymb,amsfonts}%
\usepackage{amsthm}%
\usepackage{mathrsfs}%
\usepackage[title]{appendix}%
\usepackage{xcolor}%
\usepackage{textcomp}%
\usepackage{manyfoot}%
\usepackage{booktabs}%
\usepackage{algorithm}%
\usepackage{algorithmicx}%
\usepackage{algpseudocode}%
\usepackage{listings}%

\usepackage{bm}
\usepackage{subcaption}
\usepackage{natbib}
\usepackage{soul}

\usepackage{lineno}
\begin{document}

%% ------------------------------------------------------------------------ %%
%  Title
\title[Article Title]{The effect of particle shape on sunlight exposure in a wind-driven free-surface flow}%Solar irradiance of nonspherical particles in wind-driven waves and turbulence}

%% ------------------------------------------------------------------------ %%
%
%  AUTHORS AND AFFILIATIONS
%
%% ------------------------------------------------------------------------ %%

\author*[1]{\fnm{Lucia} \sur{Baker}}\email{ljbak@uw.edu}

\author[1]{\fnm{Julio E.} \sur{Ch{\'a}vez-Dorado}}\email{jechavez@uw.edu}

\author[1]{\fnm{Anusha} \sur{Aggarwal}}\email{aa11557@uw.edu}

\author[1]{\fnm{Inessa} \sur{Garrey}}\email{ivgarrey@gmail.com}

\author[2]{\fnm{Michelle H.} \sur{DiBenedetto}}\email{mdiben@princeton.edu}

\affil*[1]{\orgdiv{Mechanical Engineering}, \orgname{University of Washington}, \orgaddress{\city{Seattle}, \postcode{98195}, \state{WA}, \country{USA}}}

\affil[2]{\orgdiv{Mechanical and Aerospace Engineering}, \orgname{Princeton University}, \orgaddress{\city{Princeton}, \postcode{08544}, \state{NJ}, \country{USA}}}

% ABSTRACT

\abstract{
We investigate how nonspherical marine particles, such as microplastics, phytoplankton, and ice, are exposed to sunlight near the water surface in wind-driven surface boundary layer conditions. We measure the depth and orientation of these particles in a laboratory wind-wave tank, then use this data to model how much sunlight the particles would be exposed to in an aquatic environment. In particular, we explore how their orientation affects their sunlight exposure in these conditions, since most studies to date do not account for orientation, turbulent shear flow, and wind-waves at the water surface. We find that in all cases, particles prefer certain orientations, and this causes the average sunlight exposure of the particles to be 5 to 40\% higher than would be expected if their orientations were random. Further, we find that the mean irradiance depends on the ratio of the particle mixing length $L_m$ to the light decay lengthscale $\lambda$, and the nature of this dependence differs between disks and rods. Disks experience disproportionately greater mean irradiance at smaller $L_m/\lambda$ due to the vertical variation in their orientations. These results have important implications for accurately estimating the sunlight exposure of particles in the surface boundary layer.
}


\maketitle

\section{Introduction}
\label{sec:intro}

The aquatic boundary layers under the surfaces of lakes, oceans, and other bodies of water are home to a variety of suspended particulates, both organic and inorganic. Many of these particulates are strongly affected by sunlight. For example, photosynthesis rates of organisms such as phytoplankton and other algae depend on solar irradiance. Microplastic particles may be photoreactive and degrade with exposure to sunlight \citep{ward2019}. The rates of melting or freezing of ice crystals or pancake ice also respond to heating from sunlight \citep{thomas2016}. 

Due to their nonspherical shape, the light exposure or irradiance of these particles depends not only on the incident sunlight, but also on their depth and orientation: sunlight intensity decays with depth in the water column, and the particle's orientation determines the normal surface area with respect to the incoming light rays. Depth and orientation are both determined by particle interactions with waves, turbulence, and wind shear in the surface boundary layer. However, research on the dynamics of aquatic microplastic particles has generally been limited to simulations which model only simplified aspects of the surface boundary layer, or studies which consider only tracer species and neglect particle shape or size \citep{dibenedetto2026}. These interactions have not yet been experimentally investigated for particles in surface boundary layer conditions.

There is a robust body of research on vertical transport of particles in a wind-driven surface boundary layer. Wind-mixing has been established as an important driver of vertical transport in lakes and the ocean and can submerge particles up to tens of meters \citep{kukulka2012,thoman2021}. Turbulence in the surface boundary layer strongly affects particle transport, as downward-directed turbulent velocity fluctuations entrain and submerge buoyant particles from the surface, while buoyancy resuspends them. At equilibrium, this process results in a vertical concentration profile that decays with depth, which depends on the balance between the entrainment and the resuspension of the particles \citep{chamecki2019}. The settling (or rise) velocity ratio $Sv^=W_t/u^*$ quantifies this balance between buoyancy and turbulent entrainment, where $W_t$ is the terminal rise velocity and $u^*$ is the characteristic shear velocity of the flow, which is related to the shear stress at the surface. 

However, spherical or tracer species are often the focus of these studies, in contrast to many realistic marine particles which have either flat (disk-like) or elongated (rod-like) shapes. Particle shape and finite size affect their transport in turbulence through their orientation-dependent drag coefficients and torques. For example, in homogeneous isotropic turbulence, phytoplankton-like elongated particles settle faster as their aspect ratio increases and cluster more than spherical particles \citep{ardekani2017}. Several studies have investigated axisymmetric non-spherical particle orientation and rotation in wall-bounded turbulent shear flows without waves; these studies have found their rotational dynamics and preferential orientations to depend in complex ways on particle shape, inertia, and aspect ratio \citep[e.g.,][]{marchioli2010,ni2015,zhao2015,challabotla2016,yuan2017,baker2022}. In particular, the particle alignment relative to the direction of the shear depends on the particle inertia and buoyancy. In the case of inertial, non-neutrally-buoyant particles, particles tend to align their longer axis with the streamwise direction of the flow. In situ measurements by \citet{nayak2018} found evidence for particle alignment in turbulent shear flows with various flow conditions for naturally-occurring particles from micron to centimeter sizes.

Fewer studies exist on particle orientation in wavy flows or free surface boundary layers. \citet{dibenedetto2018b} consider neutrally buoyant disk and rod particles in laminar, linear waves and find that the waves focus particles into a stable oscillating preferential orientation. \citet{dibenedetto2019} further find that surface gravity waves can focus non-spherical particles into preferred orientations which maximize their vertical drag coefficient, and that settling can destabilize the wave-induced preferential orientation. The combined effect of waves and shear flow on particle orientation is modeled by \citet{salemink-harry2025}, who find a 10-20\% increase in light attenuation in the water column by spheroidal particles comparable in size to algae or microplastics. This increase is attributed to preferential orientation induced by the waves and shear, which tends to increase their projected area of the particles and therefore the amount of light they intercept. However, these studies do not include the effects of turbulence. Moreover, the authors are aware of no study to date that has experimentally measured the orientation of buoyant particles in a wind-driven, free-surface boundary layer. 

To quantify how much sunlight particles receive, we consider irradiance, defined as radiant flux (or radiant power) received by a surface per unit area. Irradiance depends on light intensity, which decays with depth due to light scattering and absorption in the water column; hence, a deeper particle has a lower irradiance. Irradiance also depends on the projected surface area of the particles that can intercept sunlight, which is a function of the particle orientation. A particle whose major light-receiving surface is normal to the direction of the incoming light rays maximizes its irradiance, and its irradiance decreases as the particle tilts away from the incoming light.

Nonspherical particles' preferential orientation, therefore, sets up competing effects on their irradiance. Turbulence tends to mix particles deeper into the water column, where there is less light, so we might expect a simple relationship where turbulent conditions reduce the mean particle irradiance. Indeed, this is necessarily the case for spherical particles. However, both waves and shear flow have been shown to preferentially orient nonspherical particles in a way that increases their projected surface area. Higher windspeeds increase the strength of waves, shear, and turbulence in tandem, and the combination of these competing effects on the irradiance of nonspherical particles in a wind-driven surface boundary layer is not known.

The effect of turbulent mixing on irradiance can be quantified by considering the ratio of the particle mixing length, $L_m$, a characteristic lengthscale quantifying how deep the particles are mixed, to the light attenuation (or decay) lengthscale, $\lambda$. At the extremes, $L_m>>\lambda$ is the case where most particles are not exposed to significant amounts of light because they are mixed deeper by turbulence than most light penetrates; $L_m<<\lambda$ is the case where particles are exposed to a nearly uniform light intensity because the particles are shallow enough to be exposed to nearly unattenuated light. By controlling $L_m/\lambda$, we can isolate and explore the effects of particle orientation on irradiance. 

To investigate irradiance of marine particles in surface boundary layer conditions, we analyze the distribution of orientation and depth for inertial disk- and rod-shaped particles of various sizes in a wind-driven surface boundary layer, as well as spherical particles as controls. We use a laboratory wind-wave facility which allows us to make detailed measurements of both the fluid and particle mechanics under controlled conditions. We then apply an idealized solar irradiance model to our laboratory measurements to analyze how particle shape and the ratio of $L_m/\lambda$ affect mean particle sunlight exposure.

% ----------------- METHODS ----------------- 
\section{Methods}\label{sec:methods}

\subsection{Experimental facility and flow characteristics}\label{sec:flow}

Experiments are performed in the Washington Air-Sea Interaction Research Facility, a water channel in which wind can be blown over the surface of the water. The test section of the water channel is 12.2 m long and 0.9 m wide, and is filled with tap water to a depth of 0.6 m, leaving 0.6 m of headspace for airflow. The top of the channel is covered by removable panels except where instrumentation must pass through. Wind is generated by a suction fan that drives air through the headspace in the test section and recirculates it via an overhead duct. A sloped beach lined with coiled-fiber mats is placed at the downstream end of the tank to prevent wave reflections. The region of interest (ROI) in which measurements are taken is centered on a fetch of 6.8 m. The ROI is 1 m long in the streamwise direction and spans the full width (0.9 m) and depth (0.6 m) of the channel. For this experiment, a higher and low wind stress condition are generated by setting the windspeed to 16 m/s and 12 m/s, respectively. 

Water velocities and free surface elevations are measured using a Nortek Vectrino Profiler acoustic Doppler velocimeter (ADV) and an Akamina AWP-24-3 capacitance wave height gauge. The ADV probe, mounted on a 2D traverse, is swept through the water cross-section and records time series of 3D velocities along vertical profiles. The probe faces upward so it can also measure the distance between itself and the free surface, providing the instantaneous surface elevation associated with each velocity measurement. The wave gauge is mounted at a fixed point and is used to obtain a time series of surface elevation, from which wave characteristics are obtained. 

\begin{figure}
    \centering
    \begin{subfigure}[b]{0.5\textwidth}
        \centering
        \includegraphics[width=\textwidth]{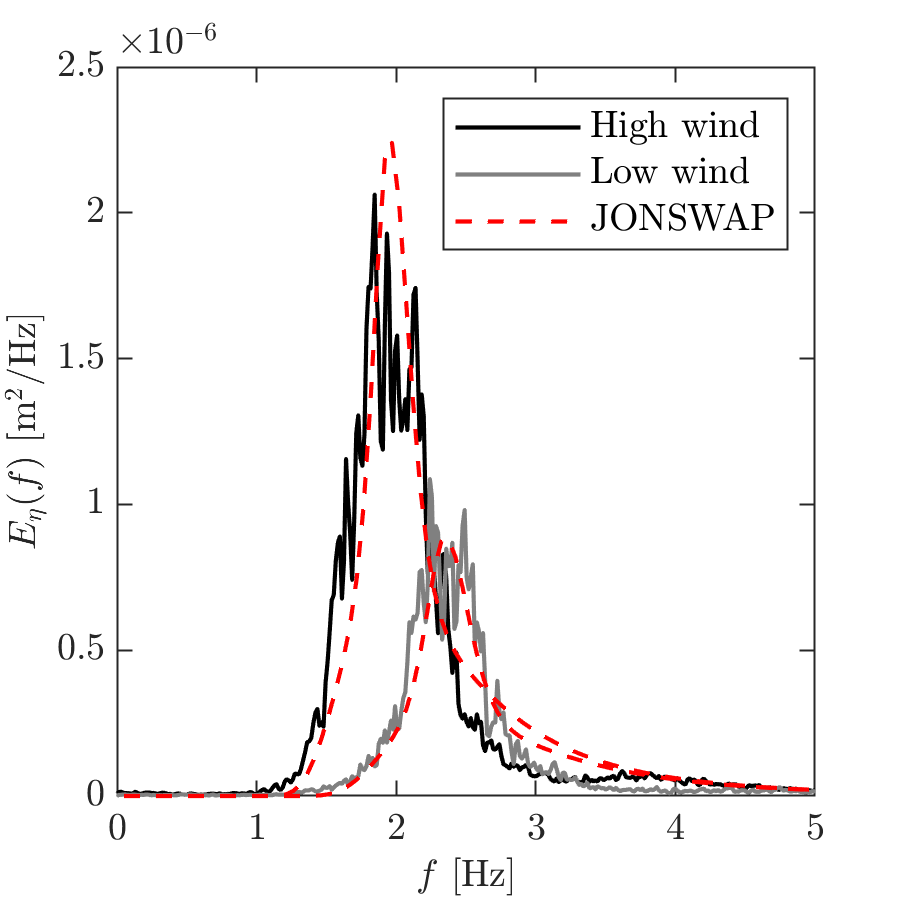}
    \end{subfigure}%
    ~
    \begin{subfigure}[b]{0.5\textwidth}
        \centering
        \includegraphics[width=\textwidth]{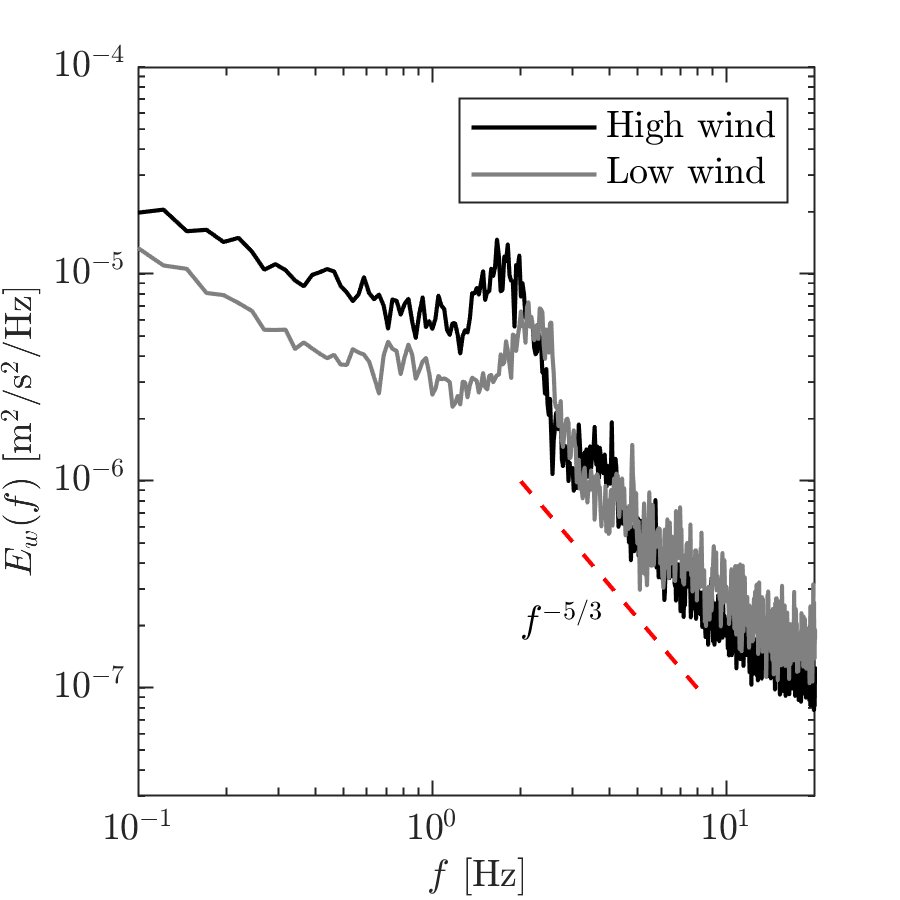}
    \end{subfigure}
    \caption{Flow spectra. (a) Wave energy spectra $E_\eta(f)$ for the high- and low-windspeed cases compared with JONSWAP spectra. (b) Energy spectra of the vertical flow velocity $E_w(f)$, showing the peak associated with the wind-waves and -5/3 power-law decay (dashed line) associated with fully turbulent flow.}\label{fig:spectra}
\end{figure}

The significant wave height $H_s$ for each windspeed case is estimated as four times the standard deviation of the surface elevation measured by the wave gauge. Wave energy spectra are computed for each case and are shown in figure \ref{fig:spectra}(a). From the peak frequency of these spectra, we estimate the dominant wavenumber $k_{dom}$ for each case. Values of significant wave height and dominant wavenumber are presented in table \ref{tab:waveflowprops}.

JONSWAP wave spectra are also fit to the data using the measured peak wave period and amplitude, and the shapes of the observed wave spectra and the JONSWAP spectra are found to match reasonably well. The JONSWAP family of spectra are derived empirically from wave measurements in the fetch-limited environment of the North Sea \citep{hasselmann1973}. The similarity between our measurements and the JONSWAP spectra demonstrates that the waves produced in our experiment, while scaled down from oceanic conditions, still have comparable distributions of wave energy to wind waves in lake and fetch-limited ocean environments.  

To characterize the flow itself, we examine the energy spectrum of the water's vertical velocity for each windspeed case, shown in figure \ref{fig:spectra}(b). Apparent in these spectra are prominent peaks corresponding to the wave oscillations. We also see a range where the energy spectrum follows a -5/3 power-law decay, which is indicative of fully turbulent flow \citep{goto2016,kolmogorov1941}. From these spectra, we demonstrate that we have a flow that is both wavy and turbulent. 

%% wave and flow properties
\begin{table}
  \centering
  \def~{\hphantom{0}}
  \begin{tabular}{cccccc}
      $U$ [m/s] & $H_s$ [cm]  &  $k_{dom}$ [m$^-1$]  &  $u^*$ [cm/s] & $Re^*$   \\ \hline
      16 & 6.3 & 15.3 & 7.4 & $4.6 \times 10^5$  \\
      12 & 4.4 & 22.2 & 4.1 & $2.5 \times 10^5$ 
  \end{tabular}
  \caption{Wave and flow properties for the two windspeeds, $U$: the significant wave height $H_s$, the dominant wavenumber $k$, the shear velocity $u^*$, and the associated Reynolds number $Re^*$.}
  \label{tab:waveflowprops}
\end{table}

This is an oscillatory flow, so we examine water velocity as a function of wave phase and depth below the free surface. In order to do this, we first convert the vertical position coordinate of our ADV measurements in the fixed laboratory reference frame, $z$, to the depth relative to the instantaneous free surface, $\tilde{z}$, which varies in time. We assume an exponential decay of wave influence with increasing depth below the surface, with the decay lengthscale equal to the $1/k_{dom}$. Thus, the surface-relative depth coordinate is given by $\tilde{z} = z - \eta\exp{(k_{dom}z)}$. We obtain the instantaneous wave phase $\phi$ by performing a Hilbert transform on the instantaneous depth-averaged streamwise velocity $\bar{u}$ as follows: $\phi = \arctan\big(\mathcal{H}(\bar{u})\big)$. Velocity data is then binned on both $\tilde{z}$ and $\phi$ to compute mean streamwise, spanwise, and vertical velocity $\langle u(\tilde{z},\phi)\rangle$, $\langle v(\tilde{z},\phi)\rangle$, and $\langle w(\tilde{z},\phi)\rangle$, respectively. Velocity fluctuations $u'(\tilde{z},\phi)$, $v'(\tilde{z},\phi)$, and $w'(\tilde{z},\phi)$ are computed by subtracting the mean velocity from each component. The water-side shear velocity $u^*$ can then be estimated by considering the shear stress at the free surface:
\begin{equation}
    \tau_s(\phi) = \rho_f\bigg(\nu \frac{\partial u}{\partial \tilde{z}} - \langle u'w'\rangle\bigg)\bigg|_{\tilde{z}=0}
\end{equation}
where $\nu$ is the kinematic viscosity of the water. Averaging over wave phase and using the definition of the shear velocity, the shear velocity is given by:
\begin{equation}
    u^* = \sqrt{\langle \tau_s(\phi)\rangle/\rho_f}
\end{equation}
The calculated values of the shear velocities for each windspeed are also used to calculate the associated Reynolds number of the flow: \begin{equation} Re^* = \frac{u^*H}{\nu} \end{equation} where $H$ is the water depth. Values of the shear velocity and Reynolds number for both windspeeds are presented in table \ref{tab:waveflowprops}. In both cases, the Reynolds number indicates that the wind-driven shear flow is fully turbulent.

\subsection{Particles and imaging}\label{sec:particles}

We then introduce particles into the flow. We use plastic rods, disks, and spheres spanning a range of sizes from mm to cm scales as our test particles. Particles are all made of high-density polyethylene (HDPE), which is slightly buoyant in water with a specific gravity of 0.95. The rods are cut from 1.75 mm thick 3D-printing filament using a razor blade, and the disks are cut from a 1.59 mm thick sheet using circular dies mounted to an arbor press. The dimensions of the particles are defined in terms of their axis lengths: $a$ is the length of the axis of rotational symmetry (i.e., the axis through the length of a rod and through the thickness of a disk), and $b$ is the length of the perpendicular axes. The long axis length (the larger of $a$ and $b$, or the diameter in the case of the spheres) is denoted by $D_p$.

In choosing the particle material and geometry, our goal is to systematically sweep over a range of parameters relevant to microplastic mixing in the surface boundary layer. The dimension $D_p$ ranges from 2.0 to 3.9 mm for the spheres, 10 to 20 mm for the rods, and 5 to 10 mm for the disks. These correspond to aspect ratios $a/b$ of 1 for the spheres, $O(10)$ for the rods, and $O(0.1)$ for the disks. The particles' terminal rise velocities $W_t$ fall between 1.3 and 3.7 cm/s, where particles with larger volumes generally have higher rise velocities. These characteristics result in particles with rise velocity ratios $Sv=W_t/u^*$ in the range 0.2 to 0.5 in the high-wind case and 0.3 to 0.9 in the low-wind case. The rise velocity ratio ranging from $O(0.1)$ to $O(1)$ simulates a large subset of marine particles with intermediate $Sv$ signifying scenarios when both particle buoyancy and turbulent entrainment are important \citep{dibenedetto2023,chor2018}. Complete particle properties are presented in table \ref{tab:dimpartprops}.

\begin{table}[]
    \centering
    \begin{tabular}{lcccccccc}%{l|c|c|c|c|c|c|c|c}
        Particle type & S2 & S3 & S4 & R10 & R20 & D5 & D7 & D10 \\ \hline
        $a$ [mm] & 2.0 & 2.5 & 3.9 & 10.7 & 20.2 & 0.79 & 0.79 & 0.79 \\
        $b$ [mm] & - & - & - & 1.8 & 1.8 & 5.0 & 7.0 & 10.0 \\
        $a/b$ & 1 & 1 & 1 & 5.7 & 11.4 & 0.16 & 0.11 & 0.08 \\
        $\rho_p/\rho_f$ & 0.95 & 0.95 & 0.95 & 0.95 & 0.95 & 0.95 & 0.95 & 0.95 \\
        $W_t$ [m/s] & 0.026 & 0.033 & 0.037 & 0.025 & 0.021 & 0.013 & 0.013 & 0.020 \\
        $Sv^*$ ($U=16$ m/s) & 0.3 & 0.4 & 0.5 & 0.3 & 0.3 & 0.2 & 0.2 & 0.3 \\
        $Sv^*$ ($U=12$ m/s) & 0.6 & 0.8 & 0.9 & 0.6 & 0.5 & 0.3 & 0.3 & 0.5
    \end{tabular}
    \caption{Physical properties of the eight types of plastic particles: the length of the rotational symmetry axis $a$, the length of the perpendicular axis $b$, the aspect ratio $a/b$, the specific gravity $\rho_p/\rho_f$, the terminal rise velocity $W_t$, and the rise velocity ratios $Sv=W_t/u^*$ for each windspeed case. Each particle type is referred to by a letter corresponding to their shape (Sphere, Rod or Disk) and a number corresponding to their approximate size in mm.}
    \label{tab:dimpartprops}
\end{table}

%\subsection{Imaging and image processing}

We measure the particle position and orientation for each windspeed and particle type using a shadow-tracking method. The method is described in detail in \citet{baker2023}. Briefly: four synchronized cameras are used to capture images of the ROI at a rate of 30 frames per second. The ROI is illuminated with collimated light, and the cameras are focused on the back wall of the water channel to capture image series of the particle shadows. The 2D positions are then determined from the centroids of the shadows, and the 3D orientations are calculated trigonometrically from measurements of the shadows' major and minor axes.

% ----------------- RESULTS ----------------- 
\section{Results}\label{sec:results}

\subsection{Depth and orientation}

\begin{figure}
    \centering
    \includegraphics[width=0.6\textwidth]{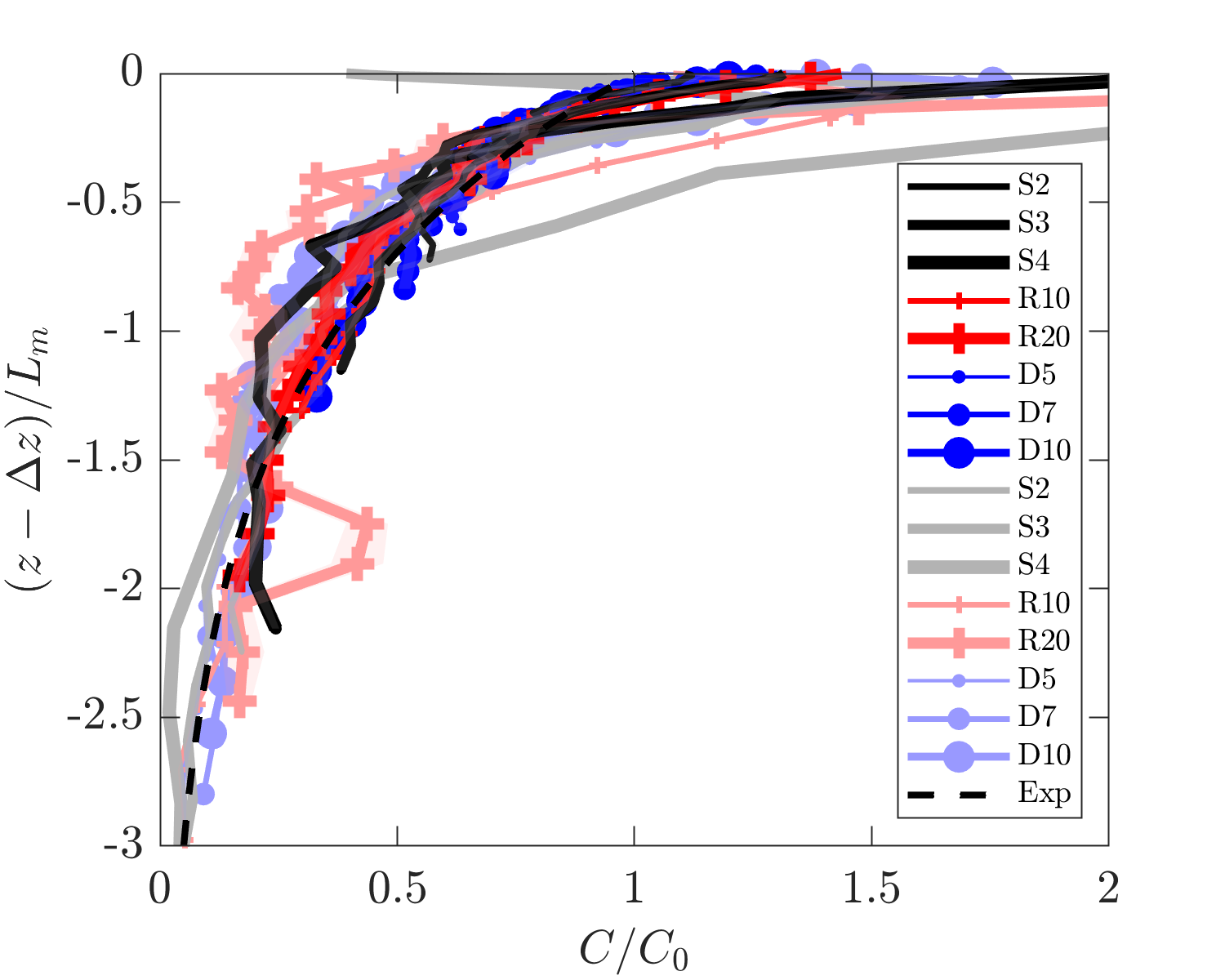}
    \caption{Profiles of concentration normalized by concentration at the surface, $C/C_0$, as a function of depth offset by $\Delta z$ and normalized by the mixing length, $(z-\Delta z)/L_m$. Profiles are shown for each particle type (rods shown in red, disks shown in blue, and particle size denoted by line thickness) and windspeed (high wind case shown in bright color, low wind case shown in light color), compared with an exponential decay (black dashed line). Here and throughout, particle types are referred to by a letter corresponding to their shape (Sphere, Rod or Disk) and a number corresponding to their approximate size in mm. Details of each particle type are provided in table \ref{tab:dimpartprops}.}
    \label{fig:conc-fit}
\end{figure}

We first examine the vertical profile of mean particle concentration as a function of depth. The observed concentration profiles are shown in figure \ref{fig:conc-fit}. Field observations in a wind-driven surface boundary layer by \citet{kukulka2012} show that concentration often follows an exponential decay with depth. This is in contrast to solid-wall turbulent boundary layers, in which concentration follows a power-law distribution with distance from the wall \citep{rouse1937}. Here, we observe that the concentration profiles, including the sphere cases, collapse on the exponential decay profile $C/C_0=\exp{\big(\frac{z-\Delta z}{L_m}\big)}$. Here, $C_0$ is the concentration at the surface, $\Delta z$ is an offset to the depth coordinate roughly equal to the significant wave height, and the decay lengthscale $L_m$ is the mixing length. Because $C_0$ and $\Delta z$ are fixed, we use a least-squares regression fit of each concentration profile to obtain $L_m$ for each particle type and windspeed case.

We then consider particle orientation. In the context of sunlight exposure, the relevant measure of orientation is the angle $\alpha$ between the direction of the incident light rays and the vector normal to the particle's major light-receiving surface. The light received by each particle shape is then proportional to $\cos \alpha$. We start by defining an orientation vector $\boldsymbol{p}$ which passes through a particle's axis of rotational symmetry, as in \citet{voth2017}, which we are able to measure experimentally. For disks, the major light-receiving surface is normal to $\boldsymbol{p}$ (i.e., the disk's face), and for rods it runs parallel to $\boldsymbol{p}$ (i.e., along the rod's length). Further, in this analysis, we assume that the light rays come from directly overhead (a zero angle of incidence). This is reasonable as a first-order approximation for two reasons: (1) if we are interested in the cumulative effects of radiation, these effects will predominantly accumulate when the radiation is strongest, i.e., around midday when the sun is at low angles of incidence and travels through less atmosphere; (2) refraction as light enters the water tends to reduce the incident angle of the light rays on underwater particles. Therefore, we calculate $\cos\alpha$ from $p_z$ for each particle shape as follows:
\begin{align}
    &\text{Rods: } \cos\alpha = \sqrt{1-p_z^2}    \\
    &\text{Disks: } \cos\alpha = \big|p_z\big|
\end{align}

\begin{figure}
    \centering
    \includegraphics[width=\textwidth]{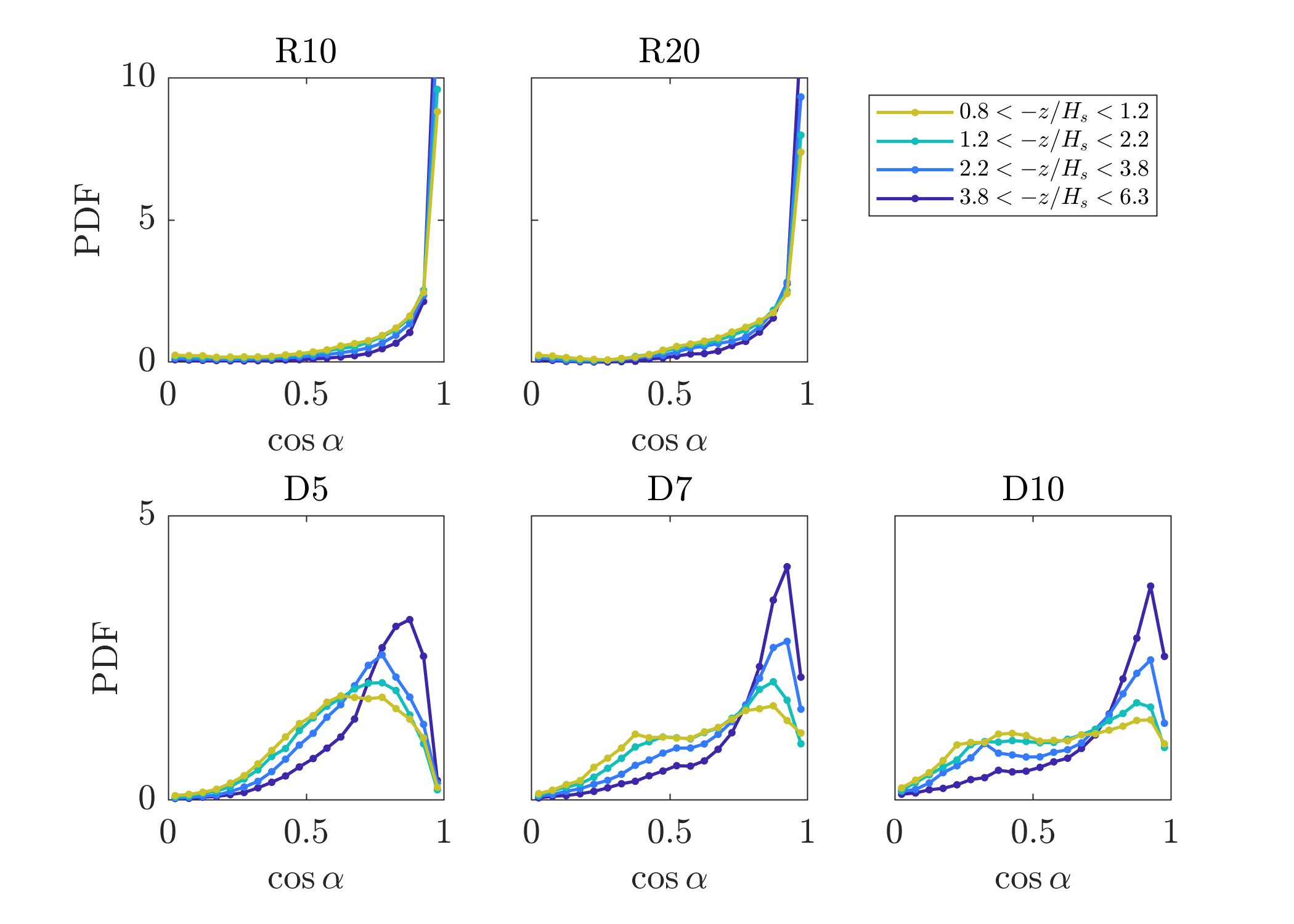}
    \caption{Probability distribution functions (PDFs) of $\cos\alpha$ binned on depth normalized by significant wave height, $z/H_s$. PDFs are computed on logarithmically-spaced depth bins, where the shallowest bin is shown in yellow and the deepest bin shown in dark blue. Only the high-wind case is shown for clarity, but the same trends hold in the low-wind case.}
    \label{fig:cosa-depth}
\end{figure}

Probability distribution functions (PDFs) of $\cos\alpha$, binned on depth, are shown in figure \ref{fig:cosa-depth}. We find evidence of preferential orientation for each particle type throughout the water column, with the distributions peaking in all cases at high $\cos\alpha$ values, showing that particles prefer to normalize their major light-receiving surfaces to the vertical. In addition, the orientation distributions have a larger spread at shallower depths, indicating more randomized orientations near the surface. The near-surface randomization is pronounced for the disks, indicating that they tend to normalize to the vertical very strongly at depth and more weakly near the surface. The effect is relatively minor for the rods, and we see that they strongly normalize to the vertical both at depth and near the surface.

The surface is where both turbulence and the wind-waves' orbital velocities are the greatest, and we see they both evidently act to randomize particle orientations. Both disk and rod particles still prefer to normalize with the vertical throughout the water column, but disks are much more sensitive to the randomizing effects near the surface. These results are consistent with observations in turbulent boundary layer flows over a solid wall \citep{baker2022}, suggesting that free surface boundary layer dynamics have similar effects on particle orientation. 

% orientation pdfs vs depth

\subsection{Irradiance}

Knowing particle depth and orientation, we are able to investigate irradiance, defined as radiant flux incident on a surface per unit area. The irradiance of a particle with depth $z$ and orientation angle $\alpha$ is given by the product of the local light intensity and the cosine of its orientation angle: 
\begin{equation} R(z, \alpha) = I(z)\cos\alpha \end{equation} 
Here, $I(z) = I_0\exp(z/\lambda)$ is the irradiance at depth $z$ in the water column, where $\lambda$ is the light decay lengthscale and $I_0$ is the irradiance at the surface. In practice, both $I_0$ and $\lambda$ vary widely: $I_0$ depends on weather, season, latitude, and transparency of the atmosphere; $\lambda$ depends on transparency of the water column. In this analysis, we only consider normalized irradiance values; therefore the value of $I_0$ is arbitrary. We vary the ratio of the particle mixing length to the light decay lengthscale $L_m/\lambda$ as an independent variable by varying $\lambda$ and using the experimentally determined value of $L_m$. We investigate intermediate values of this ratio in the range $O(0.1)$ to $O(10)$. 

We then calculate mean particle irradiance for a given $L_m/\lambda$. The mean irradiance $\bar{R}$ of a set of $N$ particle observations is given by taking the ensemble average of the irradiance of each particle observation, modeled from the particle's depth and orientation: 
\begin{equation} \bar{R} = \frac{1}{N}\sum_{i=1}^NI_i\cos\alpha_i \end{equation} 
This model assumes the ability to observe infinitely deep into the water column. However, our experimental data is truncated at the bottom of the experimental region of interest (ROI, see section \ref{sec:methods}), introducing a small amount of error. To correct for this, we extrapolate particle count and irradiance between the bottom of the ROI, $z_{min}$, and $z=-\infty$. As shown earlier, particles deep in the water column are preferentially oriented with their long axis horizontal (i.e., $\cos{\alpha} \approx 1$), so we make the approximation that all particles below the bottom of the ROI would be oriented as such. We assume an exponential decay for particle count $n(z) = n_0\exp(z/L_m)$ based on the observed concentration profiles, where $n_0$ is the particle count at the surface. The mean irradiance $\bar{R}$ is then given by: 
\begin{equation}
    \bar{R}= \frac{\displaystyle\sum_{i=1}^NI_i\cos\alpha_i + \int_{-\infty}^{z_{min}} Indz}{N + \int_{-\infty}^{z_{min}}ndz}
\end{equation}
This can be integrated analytically to yield:
\begin{equation}
    \bar{R}= \frac{\displaystyle\sum_{i=1}^NI_i\cos\alpha_i + \frac{I_0n_0\lambda L_m}{\lambda+L_m}\exp{\bigg(\frac{z_{min}(\lambda+L_m)}{\lambda L_m}\bigg)}}{N + n_0 L_m\exp{\big(\frac{z_{min}}{L_m}\big)}}
\end{equation}

\begin{figure}
    \centering
    \includegraphics[width=0.6\textwidth]{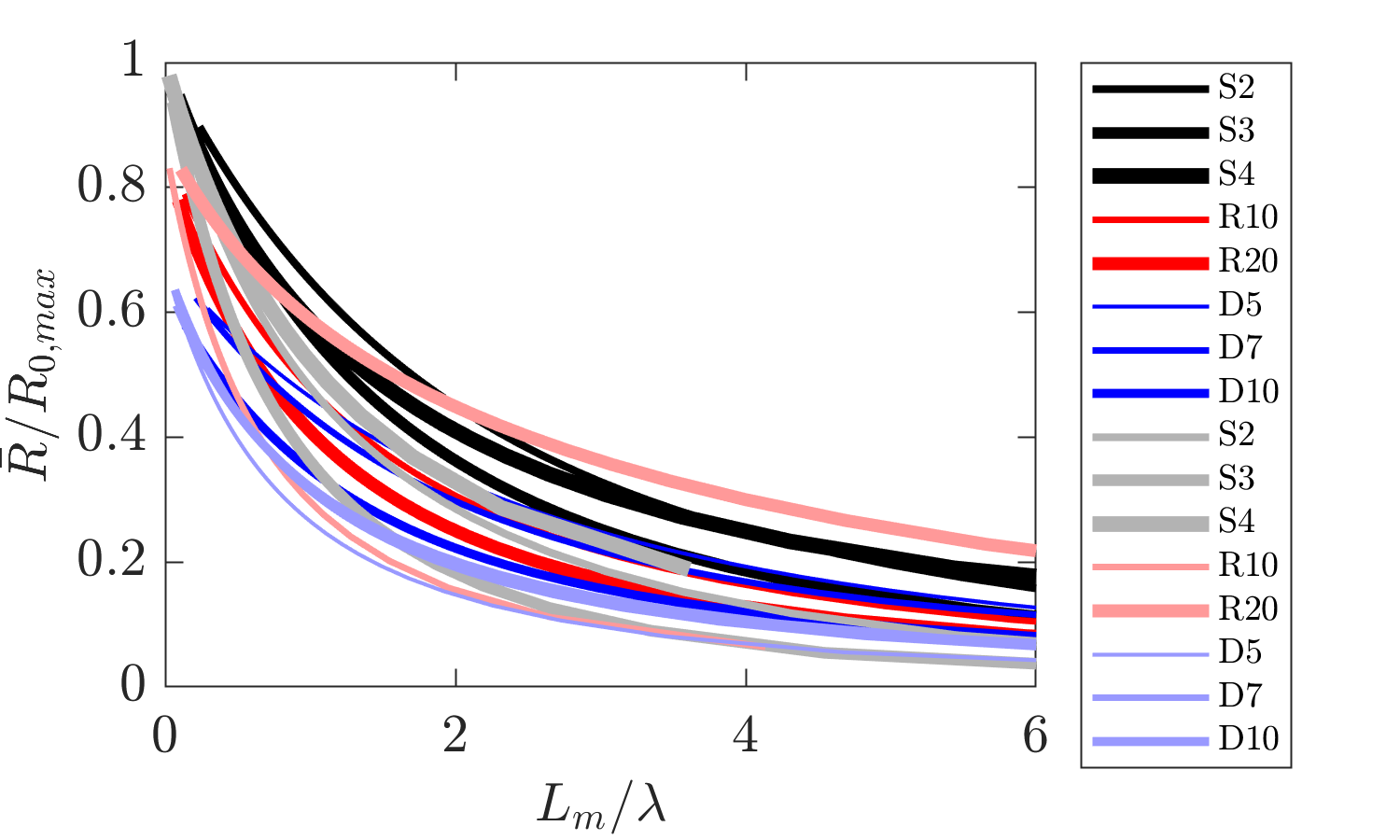}
    \caption{Mean irradiance as a function of the ratio of mixing length to light decay lengthscale, $L_m/\lambda$. The observed mean irradiance $\bar{R}$ is normalized by $R_{0,max}$, the irradiance at the surface for particles with maximized $\cos\alpha$. The normalized mean irradiance is compared for each particle type (rods shown in red, disks shown in blue, and particle size denoted by line thickness) and windspeed (high wind case shown in bright color, low wind case shown in light color).}
    \label{fig:R_R0_max}
\end{figure}

The observed mean irradiance is shown as a function of the ratio $L_m/\lambda$ in figure \ref{fig:R_R0_max}. Here, $\bar{R}$ is normalized by the maximum surface irradiance $R_{0,max}$, the irradiance of a particle at the surface ($I=I_0$) with maximized surface area normal to incoming light ($\cos\alpha=1$). We include spheres for reference, which have no orientation effect. 

In all cases, mean irradiance decreases with increasing mixing length relative to the light attenuation lengthscale. In other words, between the competing effects of turbulent mixing and preferential orientation on irradiance discussed in section \ref{sec:intro}, turbulent mixing is dominant. However, the mean irradiance curves do not fully collapse on $L_m/\lambda$, indicating that irradiance does not depend on turbulent mixing effects alone. 

\begin{figure}
    \centering
    \includegraphics[width=\textwidth]{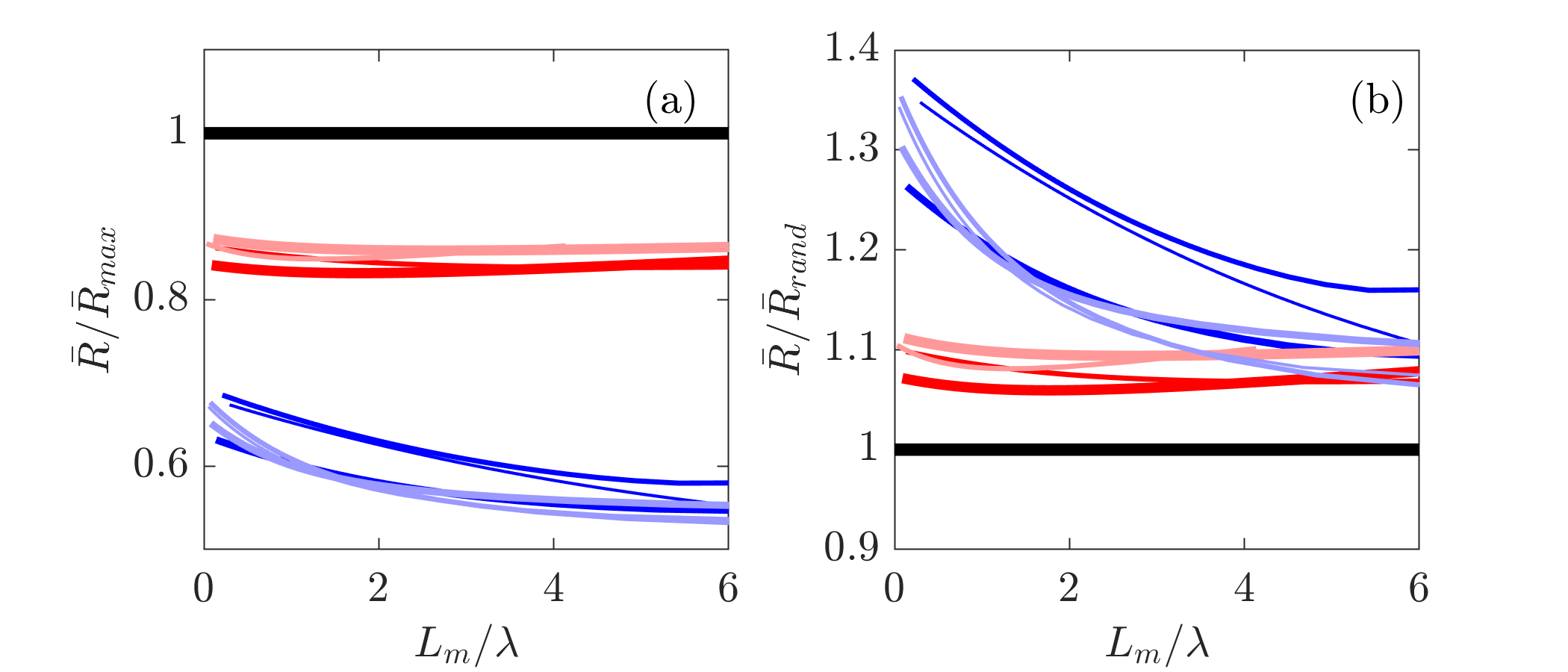}
    \caption{Mean irradiance as a function of the ratio of mixing length to light decay lengthscale, $L_m/\lambda$, normalized to isolate the effects of particle orientation. The observed mean irradiance $\bar{R}$ is normalized by: (a) $\bar{R}_{max}$, the mean irradiance for particles with perfectly maximized $\cos\alpha$, and (b) $\bar{R}_{rand}$, the mean irradiance for particles with perfectly randomized $\cos\alpha$. The black reference lines denote $\bar{R} = \bar{R}_{max}$ and $\bar{R} = \bar{R}_{rand}$, respectively. The normalized mean irradiance is shown for each particle type and windspeed (legend same as figure \ref{fig:R_R0_max}).}
    \label{fig:R_Rmax_Rrand}
\end{figure}

In order to isolate the effect of orientation, we normalize the mean irradiance $\bar{R}$ by what the mean irradiance would be if the particles were perfectly normally oriented to maximize $\cos\alpha$. This maximized irradiance, $\bar{R}_{max}$, is achieved when $\cos\alpha=1$ for all particles. Likewise, we normalize $\bar{R}$ by what the mean irradiance would be if all particles were perfectly randomly oriented. The mean $\cos\alpha$ in the randomly-oriented case can be derived analytically, and is $\cos\alpha_{rand} = \pi/4$ for rods and $\cos\alpha_{rand} = 1/2$ for disks. Hence, the mean irradiance for a perfectly maximized or perfectly randomized orientation distribution is given, respectively, by:
\begin{align}
    \bar{R}_{max} &= \bar{I}  \\
    \bar{R}_{rand} &= \bar{I}\cos\alpha_{rand}
\end{align} 
where
\begin{equation}
    \bar{I} = \frac{\displaystyle\sum_{i=1}^NI_i + \frac{I_0n_0\lambda L_m}{\lambda+L_m}\exp{\bigg(\frac{z_{min}(\lambda+L_m)}{\lambda L_m}\bigg)}}{N + n_0 L_m\exp{\big(\frac{z_{min}}{L_m}\big)}}
\end{equation} 
The observed mean irradiance normalized by what the mean irradiance would be in the maximized and randomized cases is shown as a function of the ratio $L_m/\lambda$ in figure \ref{fig:R_Rmax_Rrand}. 

We find consistent differences between the spherical and nonspherical particles, as well as between disks and rods. $\bar{R}/\bar{R}_{max}$ ranges from 0.55 to 0.7 for disks, i.e., the mean irradiance of disks is 55 to 70\% of what would be expected if their orientations maximized $\cos\alpha$, and $\bar{R}/\bar{R}_{max}$ ranges from 0.85 to 0.9 for rods (figure \ref{fig:R_Rmax_Rrand}a). This also means that mean irradiance is less for both disks and rods than for spheres for a given maximum particle cross-sectional area. There is less of a reduction for rods due to their stronger preferential orientation throughout the water column, especially near the surface, compared to disks. Because disk orientations are more randomized near the surface than rods, where the light intensity is the highest, their mean irradiance is reduced even more relative to $\bar{R}_{max}$. 

At the same time, $\bar{R}/\bar{R}_{rand}$ ranges from 1.05 to 1.4 for disks, i.e., the mean irradiance of disks is 5 to 40\% higher than would be expected for perfectly randomized orientations, and $\bar{R}/\bar{R}_{rand}$ ranges from 1.05 to 1.12 for rods (figure \ref{fig:R_Rmax_Rrand}b). This also means that mean irradiance for disks and rods is greater than for spheres for a given particle volume. We can again attribute this to the particles' preferential orientation. Because both disk and rod particles preferentially tend towards higher $\cos\alpha$, the irradiance of both shapes is greater than if they were randomly oriented.  

In addition, because particle orientation distributions vary with depth, mean irradiance also depends on the ratio $L_m/\lambda$. Disks have a higher mean irradiance for smaller $L_m/\lambda$ due to their stronger preferential orientation at depth and weaker preferential orientation near the surface. In contrast, the average irradiance of the rods is essentially independent of $L_m/\lambda$, consistent with their strong preferential alignment throughout the water column that weakens only slightly near the surface. 

\section{Discussion}
By analyzing depth and orientation of rod and disk particles in a laboratory-simulated surface boundary layer, we have estimated the relative effects of particle shape and ratio of mixing length to light decay lengthscale $L_m/\lambda$ on mean irradiance $\bar{R}$. From our laboratory measurements, we find that the concentrations of both rods and disks decay exponentially with depth (figure \ref{fig:conc-fit}), which agrees with established principles of particle concentration in wind mixing. However, we do find different orientation patterns between disks and rods (figure \ref{fig:cosa-depth}). Rods and disks are both preferentially oriented in a way that tends to normalize their surface area to incoming light. Rods are strongly preferentially oriented throughout the water column, but disks tend to be more randomized near the surface where the shear, turbulence, and waves are the strongest.

The effects of particle mixing on irradiance are shown in figure \ref{fig:R_R0_max}. In all cases, mean irradiance decreases as $L_m/\lambda$ increases, reflecting the reduction in light levels when particles mix deeper into the water column relative to the light attenuation lengthscale. However, the mean irradiance does not depend on turbulent mixing alone. The effects of orientation are isolated by normalizing mean irradiance by what it would be if all particles were perfectly normalized (figure \ref{fig:R_Rmax_Rrand}a) or perfectly randomized (figure \ref{fig:R_Rmax_Rrand}b). We find that the irradiance of nonspherical particles is lower than it would be if all particles were normalized to the incoming sunlight, but higher than it would be if they had fully randomized orientations. Clearly, it is necessary to consider how nonspherical particles actually orient in order to accurately understand their sunlight exposure.

These results have important implications for estimating solar irradiance of particles in the surface boundary layer in aquatic environments. Most estimates of irradiance on particles in the surface layer do not take realistic particle orientation into account, or assume particles are spheres. However, our results show that neglecting particle shape and orientation may result in significant errors in estimating the solar irradiance of many types of particles found near the water surface, including microplastics, phytoplankton, and ice. Our estimation of shape effects on irradiance for disk and rod particles based on controlled experimental data serves as a starting point for further analysis into the impact of particle shape and orientation on process rates in lake and ocean environments.

%%% End of body of article

%%%%%%%%%%%%%%%%%%%%%%%%%%%%%%%%%%%%%%%%%%%%%%%
\bmhead{Acknowledgements}
We thank C. Abarca for his assistance in performing the experiments.

\section*{Declarations}

\subsection*{Funding} 
This work is supported by the National Science Foundation Division of Ocean Sciences under Grant No. 2126193.

\subsection*{Competing interests} 
All authors declare no competing interests.

\subsection*{Ethics approval and consent to participate} 
N/A

\subsection*{Consent for publication} 
N/A

\subsection*{Data availability} 
The experimental data that support the results presented in this article are publicly available in Zenodo at https://doi.org/10.5281/zenodo.19008833. 

\subsection*{Materials availability} 
N/A

\subsection*{Code availability} 
N/A

\subsection*{Author contribution}
Lucia J. Baker: Conceptualization, Data Curation, Formal Analysis, Funding Acquisition, Investigation, Methodology, Project Administration, Resources, Software, Supervision, Validation, Visualization, Writing – Original Draft Preparation, Writing – Review \& Editing.
Julio E. Ch{\'a}vez-Dorado: Formal Analysis, Investigation, Software, Methodology.
Anusha Aggarwal: Formal Analysis, Investigation, Software, Methodology.
Inessa Garrey: Formal Analysis, Investigation, Methodology.
Michelle H. DiBenedetto: Conceptualization, Funding Acquisition, Methodology, Project Administration, Resources, Supervision, Writing – Review \& Editing.

%%%%%%%%%%%%%%%%%%%%%%%%%%%%%%%%%%%%%%%%%%%%%%%%%%%%%%%%%%%%%%%%
%% References and Citations

\bibliography{bib/aniso-particles,bib/microplastics,bib/particles-BL2}

\end{document}